\documentclass[final,5p,times,twocolumn]{elsarticle}

\usepackage{hyperref} 

\biboptions{sort&compress}

\journal{Materials Today Communications}

\usepackage{threeparttable}
\usepackage{amssymb}
\usepackage{amsmath}
\usepackage{tabularx}
\usepackage{booktabs} 
\usepackage{siunitx}
\DeclareSIUnit\molar{\mole\per\cubic\decimetre}
\DeclareSIUnit\Molar{\textsc{m}} 

\journal{Materials Today Communications}

\begin{document}

\begin{frontmatter}



\title{Liquid-phase encapsulation of \(\pi\)-conjugated dyes in boron nitride nanotubes: Ensemble and single-nanotube optical characterization} 

\author[aff1]{Friedrich Schöppler}
\author[aff1]{Lukas Stumpf}
\author[aff1]{Lucas Fuhl}
\author[aff1]{Elena Behr}
\author[aff4]{Charlotte Allard}
\author[aff3]{Christoph Lambert}
\author[aff2]{Richard Martel}
\author[aff1]{Tobias Hertel}

\affiliation[aff1]{organization={Institute of Physical and Theoretical Chemistry, Julius Maximilian University Würzburg},
            addressline={Am Hubland}, 
            city={Würzburg},
            postcode={97074}, 
            state={Bavaria},
            country={Germany}}

\affiliation[aff2]{organization={Département de Chimie and Institut Courtois, Université de Montréal},
            addressline={C.P. 6128 Succ. Centre-Ville}, 
            city={Montréal},
            postcode={ H3C 3J7}, 
            state={Québec},
            country={Canada}}
\affiliation[aff4]{organization={Département de Génie Physique, Polytechnique Montréal},
            addressline={5155 Chemin de la Rampe}, 
            city={Montréal},
            postcode={ H3C 3A7}, 
            state={Québec},
            country={Canada}}
\affiliation[aff3]{organization={Institute of Organic Chemistry, Julius Maximilian University Würzburg},
            addressline={Am Hubland}, 
            city={Würzburg},
            postcode={97074}, 
            state={Bavaria},
            country={Germany}}

\begin{abstract}
Boron nitride nanotubes (BNNTs) provide wide-bandgap, optically transparent one-dimensional hosts for molecular dyes, limiting direct electronic participation of the host. Whether dye@BNNT systems produce bright, well-defined J- or H-aggregates or instead heterogeneous emissive ensembles whose character depends on chain length and local packing remains only partly resolved. We address this question using ensemble extinction, photoluminescence, quantum-yield measurements, and TCSPC-derived radiative and non-radiative rates, together with polarization-resolved single-nanotube microscopy on encapsulated quaterthiophene, sexithiophene, octithiophene, and Nile Red, selected from a ten-dye screening. In the oligothiophene series, confinement modifies spectra and excited-state dynamics in a length-dependent manner, with all three oligothiophenes forming weakly emissive ensembles with suppressed effective radiative rates and 6T showing the strongest redistribution between effective radiative and non-radiative decay. The absence of radiative-rate enhancement or fluorescence-lifetime shortening across the series disfavors bright J-aggregate assignments. Polarization-resolved single-nanotube microscopy reveals strongly polarized emission, but with tube-to-tube and intratube variations, identifying oligothiophene@BNNTs as ordered yet structurally heterogeneous confined ensembles. Nile Red provides a complementary case in which the dominant response is dielectric tuning of a solvatochromic charge-transfer state rather than oligothiophene-like aggregate formation. These findings establish dye-filled BNNTs as optically quiet nanoconfined systems in which molecular ordering, dielectric confinement, and guest–guest coupling can be distinguished through combined ensemble and single-nanotube spectroscopy.
\end{abstract}


\end{frontmatter}

\section{Introduction}

Boron nitride nanotubes (BNNTs) provide a chemically inert, wide-bandgap, quasi-one-dimensional host that can confine molecular guests while remaining largely transparent throughout the visible spectral range. These properties make BNNTs attractive as dielectric nanocavities for the encapsulation and optical study of \(\pi\)-conjugated dyes, because changes in absorption and photoluminescence can then be attributed primarily to confinement, molecular organization, and guest--guest interactions rather than to strong electronic coupling to the host \cite{Allard2020,Badon2023,Juergensen2025,Saeki2025}. In this respect, BNNTs differ fundamentally from carbon nanotubes, where the electronically active host framework can couple much more directly to encapsulated dyes \cite{Yanagi2007, Gaufres2016, Cadena2021, Forel2022}. They also offer advantages relative to other one-dimensional host systems such as zeolite channels, in which nanoconfinement can likewise impose dye orientation, promote anisotropic emission, and support energy-transfer cascades, but where channels are embedded in strongly scattering three-dimensional nanocrystals, making direct optical access to an individual host structure practically impossible \cite{Calzaferri2003}. By contrast, BNNTs combine optical transparency with nanoscale isolation of individual host structures, enabling direct comparison of ensemble and single-nanotube optical response.

Encapsulation of organic dyes in BNNTs has already revealed that one-dimensional confinement can produce pronounced spectral shifts, modified vibronic structure, and strong polarization anisotropy \cite{Allard2020,Badon2023,Juergensen2025,Saeki2025}. In a pioneering study, Allard \textit{et al.} showed that encapsulation of \(\alpha\)-sexithiophene (6T) and diketopyrrolopyrrole derivatives in BNNTs induces marked bathochromic shifts and substantial changes in absorption and emission spectra relative to dilute solution \cite{Allard2020}. Subsequent structural and polarization-resolved studies demonstrated that encapsulated 6T can align strongly along the nanotube axis and that the resulting aggregate geometry depends on nanotube diameter, ranging from more ordered single-file arrangements in narrow tubes to more complex multifile packings in wider cavities \cite{Badon2023}. More recent work has refined this picture further by correlating optical response with packing dimensionality and excitonic interactions in confined 6T assemblies \cite{Juergensen2025,Saeki2025}. Other studies have begun to address the dynamic behavior of encapsulated assemblies, including curvature-driven activated diffusion of one-dimensional J-aggregates inside BNNTs \cite{Marceau2025}. Together, these studies establish BNNTs as more than passive containers: they can act as one-dimensional templates that reorganize molecular dyes into optically distinct confined assemblies.

Beyond optical dye encapsulation, BNNT filling has also been explored for a wider variety of guests, including inorganic species \cite{Milligan2024}, tellurium nanowires, polyoxometalates, and proposed drug-delivery payloads, underlining the general versatility of BNNTs as host structures \cite{Qin2020,Lee2020,ElKhalifi2016,Nejad2020,Das2023}. For optical host--guest systems, however, systematic comparative studies across chemically distinct dye classes remain scarce. In particular, reliable wet-chemical filling of multi-walled BNNTs remains only incompletely understood, and unsuccessful encapsulation attempts are rarely documented explicitly.

Several practical and mechanistic questions therefore remain open. First, the selectivity of wet-chemical BNNT filling across different dye classes is still poorly mapped under comparable preparation conditions. Second, although BNNTs are particularly well suited for combining ensemble and single-particle optical measurements, most previous studies have focused on spectral shifts and structural assignments in a small number of model systems, whereas broader comparisons of absorption, photoluminescence, excited-state dynamics, and single-particle optical anisotropy remain limited. Third, it is still not well understood to what extent filling selectivity is governed by simple steric accessibility as opposed to less obvious factors such as dye solubility, chemical stability, aggregation tendency, and packing under confinement. These questions are especially relevant for multi-walled BNNT samples, which exhibit broad diameter distributions and can therefore support multiple endohedral packing motifs and substantial structural heterogeneity.

In this work, we take a comparative approach to wet-chemical dye encapsulation in multi-walled BNNTs by applying a common filling protocol to ten dyes, comprising the oligothiophenes quaterthiophene (4T), sexithiophene (6T), and octithiophene (8T), the phenoxazines Nile Red and Nile Blue, the rhodamines Rhodamine B and Rhodamine 6G, and three squaraines. By combining dye screening with ensemble absorption and photoluminescence spectroscopy, photoluminescence quantum-yield and time-resolved fluorescence measurements, and polarization-resolved single-particle microscopy, we assess both the selectivity of the filling protocol and the diversity of photophysical responses that emerge under one-dimensional confinement. Under the conditions employed here, the clearest evidence of successful encapsulation is obtained for 4T, 6T, 8T, and Nile Red, whereas Nile Blue, the rhodamines, and the squaraines do not yield convincing evidence of stable encapsulation under the present protocol. This provides an experimental basis for distinguishing between successful and unsuccessful guest systems and for evaluating how BNNT confinement modifies molecular organization and optical behavior across different dye classes.

\section{Methods}
\label{sec:methods}

\subsection{Chemicals and dye selection}
Multi-walled boron nitride nanotubes (BNNTs; nominal outer diameter \(\varnothing = \qty{5 \pm 2}{\nano\meter}\)) were obtained from NRC-produced material.\cite{Kim14,MartinezRubi15} BNNT source material and its opening/purification procedure have been described previously.\cite{Allard2020, Badon2023} The dye encapsulation protocols used here were adapted from these reports.

\begin{figure*}[ht]
    \centering
    \includegraphics[width=0.97\textwidth]{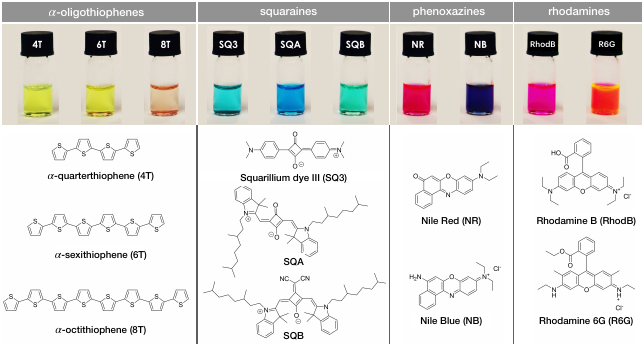}
    \caption{Overview of the molecular systems investigated in this study. Top: photographs of representative dye solutions in DMF, illustrating the distinct absorption/emission characteristics of the different dye classes. Bottom: corresponding chemical structures of selected compounds, including $\alpha$-oligothiophenes (4T, 6T, 8T), squaraine dyes (SQ3, SQA, SQB), phenoxazines (Nile Red, Nile Blue), and rhodamines (Rhodamine B, Rhodamine 6G). This set of dyes spans a range of conjugation lengths and chemical functionalities, providing a basis for comparative encapsulation and optical studies in BNNT hosts.}
    \label{fig:comparison_wide}
\end{figure*}

The dye screening comprised oligothiophenes \(\alpha\)-quaterthiophene (4T), \(\alpha\)-sexithiophene (6T), and \(\alpha\)-octithiophene (8T), the phenoxazines Nile Red (NR) and Nile Blue (NB), the rhodamines Rhodamine B (RhodB) and Rhodamine 6G (R6G), and three squaraine-type dyes (SQ3, SQA, SQB) (see Figure~\ref{fig:comparison_wide}). The oligothiophenes represent a homologous series of \(\pi\)-conjugated rod-like molecules with increasing molecular length, which makes them particularly suitable for probing size-dependent confinement and aggregation effects inside the BNNT cavity. In contrast, the remaining dyes provide more structurally diverse chromophores with different degrees of planarity, charge distribution, and solubility, allowing a broader qualitative assessment of encapsulation behavior across dye classes.

All reagents and solvents were used as received. Solvents employed during filling and washing included \textit{N},\textit{N}-dimethylformamide (DMF), toluene, dichloromethane (DCM), ethanol, and methanol. The general wet-chemical filling strategy followed literature procedures developed for dye encapsulation in BNNTs, with small adjustments for the individual dye classes \cite{Allard2020,Badon2023}.

\subsection{General wet-chemical encapsulation procedure}
\label{subsec:general_filling_procedure}
In a typical experiment, \qty{5}{\milli\gram} of BNNTs and the desired amount of dye were weighed into separate \qty{50}{\milli\liter} polypropylene centrifuge tubes. The BNNTs were dispersed in \qty{50}{\milli\liter} of solvent A, and the dye was dissolved or suspended in \qty{50}{\milli\liter} of solvent B. Each suspension was sonicated for at least \qty{15}{\minute} in an ultrasonic bath at a nominal power of about \(\qty{250}{\watt}\) at room temperature. The two liquids were then combined in a \qty{100}{\milli\liter} round-bottom flask equipped with a magnetic stir bar and sonicated again for \qty{15}{\minute} to improve mixing of partially soluble components.

The flask was fitted with a Dimroth condenser and heated under vigorous stirring to reflux for \qty{24}{\hour}, unless noted otherwise (step~E in Table~\ref{tab:filling_parameters}). After cooling, the suspension was sonicated for another \qty{15}{\minute} and filtered under reduced pressure through a PTFE membrane filter (\(\qty{0.2}{\micro\meter}\)). To remove non-encapsulated dye, the filter cake was redispersed in \qty{100}{\milli\liter} of solvent C, sonicated, and filtered again. Solvent C was selected for its particularly good solubility for the respective dye in order to wash out residual free dye from the filter cake. This washing cycle was repeated until the filtrate became colorless. In selected cases, the washing solvent was changed in the later cycles (solvent D). The final solid was dried overnight in air and subsequently redispersed in DMF for spectroscopy or spin-coated onto functionalized Si substrates for single-object microscopy. This overall protocol was adapted from the approaches reported by Allard \textit{et al.} and Badon \textit{et al.}\cite{Allard2020,Badon2023}

\subsection{Criteria used to classify encapsulation outcome}
Encapsulation was considered successful or at least plausible only when dye-specific optical signatures remained after exhaustive washing and differed reproducibly from those of the free dye in a manner consistent either with nanoscale confinement, aggregation, or a changed local dielectric environment inside the nanotube interior. Depending on the dye system, such evidence could include reproducible shifts in absorption or photoluminescence bands, changes in vibronic structure, or persistent emission signals not attributable to residual free dye.

Polarization-resolved photoluminescence measurements on individual objects were used as a complementary indicator. Strongly polarized emission from elongated emitters was taken as evidence for emission from dyes aligned within the one-dimensional nanotube cavity and thus consistent with endohedral encapsulation. By contrast, weakly polarized or unpolarized emission observed for other objects of the same sample was attributed to heterogeneity in dye orientation, loading, or local environment.

By contrast, systems were classified as unsuccessful when one or more of the following observations dominated: disappearance of the intact-dye absorption accompanied by new UV bands in the filtrate, strong solution color changes during reflux consistent with decomposition, absence of any reproducible absorption or photoluminescence signal after washing, or temporal behavior suggestive of exohedral adsorption rather than stable encapsulation.
\begin{table}[ht]
    \centering
    \caption{Parameter sets used for the filling process. The initial dye concentration $c_0$ refers to \SI{100}{mL} of solvent in the round-bottom flask. $T$ is the temperature in step E of the protocol. Solvent sequences are given as A/B/C/D, where A and B denote the initial BNNT and dye dispersion solvents, respectively, C the primary washing solvent, and D an optional secondary washing solvent. ``Tol'' = toluene.}
    
    \label{tab:filling_parameters}
    \begin{tabular}{l c l c}
    \toprule
    \textbf{Dye} & \textbf{$c_0$} / $\si{\micro mol\per\liter}$& \textbf{Solvent A/B/C/D} & \textbf{$T$ / \si{\celsius}} \\
    \midrule
    4T    & 61    & DMF/Tol/Tol/DMF\textsuperscript{a} & 120 \\
    6T    & 61    & DMF/Tol/Tol/DMF\textsuperscript{a} & 120 \\
    8T    & 63    & DMF/Tol/Tol/DMF\textsuperscript{a} & 120 \\
    \addlinespace
    SQ3   & 94    & DMF/DMF/DMF/--  & 160 \\
    SQ3   & $1900$ & DCM/DCM/DCM/DMF & 50  \\
    SQ3   & 310   & DMF/DMF/DMF/--  & 160 \\
    SQA   & 90    & DMF/DMF/DMF/--  & 160 \\
    SQB   & 90    & DMF/DMF/DMF/--  & 160 \\
    \addlinespace
    NR    & 94    & DMF/DMF/Tol/DMF & 160 \\
    NB    & 95    & DMF/DMF/EtOH/MeOH & 160 \\
    \addlinespace
    RhodB & 94    & DMF/DMF/EtOH/DMF & 160 \\
    R6G   & 94    & DMF/DMF/EtOH/DMF & 160 \\
    \bottomrule
    \end{tabular}
\vspace{1mm}
\begin{minipage}{0.98\linewidth}
\footnotesize
\textsuperscript{a} For 4T, 6T, and 8T, one additional DCM washing step was introduced after the initial DMF washing cycles, followed by additional DMF washing (4T: 1×, 6T: 4×, 8T: 1×).
\end{minipage}
\end{table}

\subsection{Successful encapsulation protocols}
\label{subsec:successful_encapsulation}

\subsubsection*{Oligothiophenes: 4T, 6T, and 8T}
Because 6T and 8T are only sparingly soluble in DMF, oligothiophenes were processed in a mixed-solvent protocol using DMF for BNNT dispersion and toluene for dye dissolution or suspension, corresponding overall to a \(1{:}1\) DMF/toluene mixture after combination. For 4T, 6T, and 8T, the nominal initial dye concentrations in the \qty{100}{\milli\liter} reaction volume were \qty{61}{\micro\mole\per\liter}, \qty{61}{\micro\mole\per\liter}, and \qty{63}{\micro\mole\per\liter}, respectively. Reflux was performed at \qty{120}{\celsius}. The solvent sequence used for filling and washing is summarized in Table~\ref{tab:filling_parameters}. For 4T and 6T, the reflux step was extended to \(3 \times \qty{8}{\hour}\). In addition, one DCM wash was introduced during the workup of all three oligothiophene samples, followed by further DMF washing as required.

These conditions yielded convincing evidence for encapsulation in the cases of 6T and 8T, and weaker but still suggestive evidence in the case of 4T. For 6T@BNNT, a pronounced red shift of the absorption maximum together with the appearance of vibronic structure was observed, consistent with earlier reports on encapsulated sexithiophene in BNNTs \cite{Allard2020}. 8T@BNNT exhibited the most pronounced spectral changes after washing relative to the free dye solutions, in both extinction and emission. 4T@BNNT showed no clearly resolvable absorption signal against the BNNT scattering background, but part of its photoluminescence was strongly red-shifted relative to free 4T after the washing procedure. The assignment of all three thiophenes as encapsulated was therefore based on spectroscopic changes relative to the free dyes after repeated washing and was later corroborated by the observation of highly polarized emission from single-particle PL microscopy in all three cases.

\subsubsection*{Nile Red}
For NR, both BNNTs and dye were processed in DMF. The nominal initial dye concentration in the \qty{100}{\milli\liter} filling mixture was \qty{94}{\micro\mole\per\liter}, and reflux was carried out at \qty{160}{\celsius}. The solvent sequence used during filling and washing is listed in Table~\ref{tab:filling_parameters}. After workup, the resulting material showed a photoluminescence band attributable to Nile Red-derived species, whereas no reliable absorption band could be extracted from the dispersion above the BNNT scattering background. The NR system was therefore retained as a promising but possibly low-loading case and characterized further spectroscopically.

\subsection{Unsuccessful encapsulation protocols}
\label{subsec:unsuccessful_encapsulation}

\subsubsection*{Nile Blue}
For NB, both BNNTs and dye were dispersed in DMF at a nominal initial dye concentration of \qty{95}{\micro\mole\per\liter}. The mixture was heated to reflux at \qty{160}{\celsius}. The solvent sequence used during filling and washing is summarized in Table~\ref{tab:filling_parameters}. Despite this workup, the resulting material did not show a reliable absorption signature of an encapsulated dye species, and the residual PL signal was weak and poorly defined. These experiments were therefore classified as unsuccessful with respect to spectroscopically validated encapsulation.

\subsubsection*{Rhodamines: Rhodamine B and Rhodamine 6G}
RhodB and R6G were both tested using DMF as solvent A and solvent B, at nominal initial dye concentrations of \qty{94}{\micro\mole\per\liter} in each case. Reflux was carried out at \qty{160}{\celsius}, and the corresponding solvent sequences for filling and washing are again listed in Table~\ref{tab:filling_parameters}. In both cases, clear changes in solution color occurred during the filling experiment, consistent with thermal and/or photochemical degradation under the reaction conditions. For RhodB, the reflux mixture turned yellow. For R6G, a red-orange coloration was observed. After filtration and repeated washing, no convincing absorption features attributable to filled BNNTs were detected. These systems were therefore not pursued further.

\subsubsection*{Squaraines: SQ3, SQA, and SQB}
Three squaraine-type dyes were examined (SQ3, SQA, and SQB). In the initial SQ3 experiment, BNNTs and SQ3 were both handled in DMF at a nominal dye concentration of \qty{94}{\micro\mole\per\liter}, followed by reflux at \qty{160}{\celsius}. The solvent sequence used in this experiment is given in Table~\ref{tab:filling_parameters}. During processing, the solution changed from turquoise to yellow. UV--vis analysis of the filtrate after the first filtration step no longer showed the characteristic absorption of intact SQ3 at long wavelength, but instead intense bands in the UV, indicating dye decomposition. A weak red-shifted feature tentatively assignable to a SQ3@BNNT species became discernible only after strong rescaling of the spectra, but the overall outcome was dominated by decomposition.

Because dilute SQ3 solutions in DMF are known to undergo light-assisted decomposition,\cite{KaczmarekKedziera2016} a second SQ3 protocol was attempted under exclusion of light in DCM at much higher concentration (\qty{1.9e3}{\micro\mole\per\liter}) and lower temperature (\qty{50}{\celsius}). The corresponding solvent sequence is likewise listed in Table~\ref{tab:filling_parameters}. Under these conditions the original dye color was preserved during heating, but the sample required extremely extensive washing (\(105\) washing cycles), and the apparent absorption signal of the product increased with time after preparation rather than remaining constant. This behavior is inconsistent with a stable encapsulated species and is more plausibly explained by dye adsorption on the outer BNNT surface followed by slow redissolution. A third SQ3 experiment was conducted again in DMF under exclusion of light at an intermediate concentration (\qty{310}{\micro\mole\per\liter}). The corresponding solvent sequence is included in Table~\ref{tab:filling_parameters}. This experiment also resulted in decomposition. Although successful squaraine encapsulation has been demonstrated previously in carbon nanotubes,\cite{Yanagi2007, Forel2022} the present BNNT experiments did not provide defensible evidence for stable SQ3 encapsulation.

SQA and SQB were processed analogously in DMF using nominal initial dye concentrations of \qty{90}{\micro\mole\per\liter} for both dyes and reflux at \qty{160}{\celsius}. In both cases, the reaction mixtures likewise changed to yellow during overnight heating. The filtrates showed no intact-dye absorption features, and the final nanotube materials displayed neither a convincing absorption nor a photoluminescence signal attributable to filled BNNTs. These systems were therefore classified as unsuccessful.

\subsection{Absorption and fluorescence spectroscopy}
UV--vis--NIR absorption spectra were recorded on dye solutions and Dye@BNNT dispersions in DMF using standard \qty{170}{\micro\liter} microcuvettes over the range \qtyrange{300}{800}{\nano\meter}. For BNNT dispersions, the measured spectra contained a significant short-wavelength scattering contribution. To minimize this effect, spectral regions containing dye absorption were masked, the residual scattering background was fitted by least squares with a double-exponential function, and the fit was subtracted from the raw spectrum. This empirical correction was used only to facilitate comparison of dye-related absorption features in colloidal dispersions. It does not alter the qualitative conclusions regarding the presence or absence of encapsulation.

Steady-state photoluminescence and excitation spectra were measured on the same cuvette format using standard fluorimeter settings. Because the spectroscopic methods were conventional and served mainly as readout tools, only the essential acquisition parameters were controlled: solvent blanks were used for background correction, excitation wavelengths were selected close to the relevant absorption bands, and detector gain was adjusted individually for each sample.

\subsection{Single-object photoluminescence microscopy and TCSPC}
For single-object studies, Dye@BNNT dispersions in DMF were spin-coated onto pre-cleaned Si substrates. Prior to deposition, the substrates were treated with piranha solution and functionalized with a 3-aminopropyltriethoxysilane (APTES) adhesion layer by vapor-phase silanization under vacuum, followed by thermal curing at \qty{100}{\celsius} to form a closed surface coating. Because the adhesion-promoting effect decreases with time, the substrates were coated with the dispersion immediately after completion of the silanization procedure. Subsequently, \qty{30}{\micro\liter} of the Dye@BNNT dispersion in DMF was spin-coated at \qty{3000}{rpm} for \qty{30}{\second}.

Polarization-resolved photoluminescence microscopy and single-object spectra were recorded on a home-built microscope based on a supercontinuum excitation source and a spectrograph/CCD detection path. The setup and sample-preparation strategy followed earlier developments from Allard et al. and our laboratory \cite{Allard2020,Fuhl2024,Oberndorfer2022}.

\subsection{Time-correlated single-photon counting}
Time-correlated single-photon counting (TCSPC) measurements were performed on selected single objects on Si substrates and, for comparison, on dye solutions stepwise diluted down to the $\mu$M range. The concentration dependence of the fluorescence decays was examined to exclude aggregation-induced effects. TCSPC was carried out on a home-built microscope using pulsed excitation from an NKT SuperK Extreme source at 485 nm, an MPD avalanche photodiode (PDM series) for detection, and a HydraHarp 400 module (PicoQuant) for photon timing. The detection path comprised a 550 nm dichroic mirror and long-pass filter together with an additional 550/750 nm bandpass combination to suppress scattered excitation light. For lifetime measurements, the emission was detected under magic-angle conditions to minimize contributions from rotational or orientational anisotropy to the measured fluorescence decays. Photon arrival times were recorded with 24 ps binning. The instrument response function had a full width at half maximum of approximately 90 ps. The count rate was kept below $1\,\%$ of the laser repetition rate to avoid pile-up effects. Measurements at different repetition rates confirmed that the full decay dynamics were captured under the chosen acquisition conditions, and the decay curves were analyzed phenomenologically using multiexponential fitting.

\subsection{Photoluminescence quantum yield measurements}
Absolute photoluminescence quantum yields were determined at room temperature using an integrating sphere (Thorlabs IS200) coupled to a fluorimeter/spectrometer setup. Measurements were carried out for free dyes in DMF and for Dye@BNNT dispersions under otherwise comparable instrumental conditions.

Samples were measured in \qty{225}{\micro\liter} quartz cuvettes (Starna 3-3.30/SOG/3). The excitation wavelengths were chosen individually for the free dyes as \qty{460}{\nano\meter} for 4T, \qty{470}{\nano\meter} for 6T, and \qty{490}{\nano\meter} for 8T. All Dye@BNNT dispersions were excited at \qty{490}{\nano\meter}. These wavelengths were selected to provide efficient excitation while maintaining consistent measurement conditions across the BNNT-containing samples. Emission was recorded over \qtyrange{510}{780}{\nano\meter} with a spectral resolution of $\approx\qty{1}{\nano\meter}$.

The quantum yields were determined by an absolute integrating-sphere method. The number of absorbed photons was obtained from the reduction of the excitation signal relative to an appropriate reference measurement, and the number of emitted photons was obtained by integration of the corresponding photoluminescence spectrum. The data were corrected for the reflectivity of the integrating sphere as well as for the wavelength-dependent response of the spectrometer and detector using the available instrumental calibration curves. Small variations in excitation power were monitored independently with a Thorlabs power meter and were taken into account in the analysis.

Each reported PL quantum yield represents the mean of repeated measurements. Based on the observed reproducibility, an uncertainty of typically \(\pm 0.01\) is assigned to the reported values. The procedure was checked against a reference measurement of Rhodamine~6G in ethanol, for which a fluorescence quantum yield of \(\Phi_\mathrm{PL}=0.92(0.05)\) has been reported in the literature \cite{Wurth2012}.

\section{Results and Discussion}
The following discussion focuses on those Dye@BNNT systems for which successful encapsulation has been established in the Methods section~\ref{sec:methods} and examines how confinement inside BNNTs modifies their optical and excited-state properties. We begin by outlining the geometric constraints for encapsulation as a useful baseline. The main focus, however, is on the spectroscopic consequences of confinement as revealed by ensemble extinction, photoluminescence (PL), quantum yield (QY), and time-resolved fluorescence (TCSPC), as well as by polarization-resolved single-particle PL microscopy. 

\subsection{Geometric constraints for dye encapsulation in BNNTs}

\begin{figure}[htbp]
    \centering
    \includegraphics[width=0.9\columnwidth]{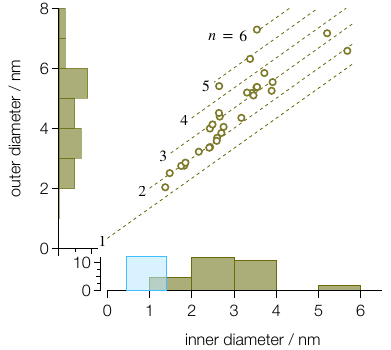}
    \caption{Correlation between inner and outer diameters of BNNTs determined from HRTEM analysis for outer diameters up to about 7\,nm, along with histograms of the inner and outer diameter distributions. Data from Allard  et al.\cite{Allard2021} Dashed sloped lines serve as tentative guides indicating discrete increments of \(\sim 0.34\,\mathrm{nm}\), corresponding to the thickness of a single BN layer. The blue square along the bottom axis indicates the estimated range of vdW dimensions of the investigated dye molecules.}
    \label{fig:diam_stats}
\end{figure}

The inner and outer diameters of the utilized BNNTs were evaluated by high-resolution transmission electron microscopy (HRTEM) by Allard et al. ,\cite{Allard2021} who assessed the range of accessible cavity sizes for dye encapsulation (Figure~\ref{fig:diam_stats}). A linear correlation between inner and outer diameter is observed, broadly consistent with nearly constant diameter offsets corresponding to even integer multiples of the BN layer spacing (\(\sim\qty{0.335}{\nano\meter}\)). The analyzed subset spans inner diameters up to nearly \qty{6}{\nano\meter}, with a broad distribution toward smaller diameters. With one exception, all observed tubes are double- or multi-walled, as indicated by the correlation between inner and outer diameter and by their alignment with the \(n \geq 2\) wall guides in Figure~\ref{fig:diam_stats}.

Following a diameter-based analysis conceptually similar to one used for the formation of carbon nanotube rings,\cite{Martel99} we note that only one BNNT in the analyzed subset is single-walled, at a diameter of about \qty{1.4}{\nano\meter}. The thermodynamically expected onset of double-wall formation \(d_\mathrm{crit}\) can be estimated from a simple balance between the curvature enthalpy of a newly formed second wall and the interwall adhesion gained upon its formation. For a cylindrical sheet bent into a cylinder of diameter \(d\), the curvature contribution per unit tube length \(L\) is expected to scale as \(H_\mathrm{curv}/L \approx 2\pi D/d\), where \(D\) is the bending rigidity of monolayer h-BN \cite{Singh2013,Sevik2011}. The adhesion gain scales as \(H_\mathrm{adh}/L \approx \pi d\,\gamma\), where \(\gamma\) denotes the h-BN/h-BN interfacial binding energy density \cite{Bjorkman2012,Leven2014}. Equating both contributions defines a critical diameter, \(d_\mathrm{crit} \approx \sqrt{2D/\gamma}\), beyond which the enthalpic balance favors formation of a second wall. Using literature values for \(D\) and \(\gamma\), this estimate places the onset on the nanometer scale, of order \(1\)--\(2\)\,nm, consistent with the observation that only a single tube near \(\sim\qty{1.4}{\nano\meter}\) appears single-walled, whereas larger tubes are double- or multi-walled.

To assess whether the observed inner-diameter distribution is compatible with dye encapsulation, Figure~\ref{fig:diam_stats} compares the BNNT inner diameters with the approximate minimal van der Waals (vdW) cross-sectional dimensions of the investigated dyes, summarized in Table~\ref{tab:vdw_dimensions_full}. Encapsulation requires that there exists at least one molecular orientation for which the full vdW cross section fits within the inner diameter of the nanotube. For anisotropic \(\pi\)-conjugated molecules, this is typically achieved by approximate alignment of the molecular backbone along the tube axis. The resulting minimal transverse dimensions are about \(0.45\)–\(0.50\)\,nm for the oligothiophenes, \(0.6\)–\(0.9\)\,nm for Nile Red and Nile Blue, up to \(0.9\)–\(1.2\)\,nm for the rhodamines, and \(\lesssim 1.4\)\,nm for the squaraines. By comparison, the BNNTs used here have an average inner diameter of about \qty{3.0}{\nano\meter}. Even allowing for a steric exclusion layer of roughly \(0.3\)–\(0.4\)\,nm at the tube wall, the remaining free diameter is still substantially larger than the minimal projected cross sections of all investigated dyes. 

Geometric exclusion alone is therefore unlikely to be the main factor limiting encapsulation under the present conditions. More relevant constraints are likely to arise from the balance of dye--wall, dye--solvent, and dye--dye interactions, together with conformational and entropic effects associated with insertion and confinement. Polarity, dispersive affinity to the BNNT interior, and the tendency of a dye to remain solvated or to aggregate may therefore influence encapsulation more strongly than simple steric fit. Although the present estimates are intentionally approximate and do not account for solvent-mediated insertion pathways or kinetic limitations of the filling process, they provide a useful geometric baseline for interpreting the experimentally observed trends in filling efficiency across the different dye classes.

\begin{table*}[ht]
    \centering
    \caption{Approximate van der Waals (vdW) cross-sectional dimensions of the investigated dyes. $d_\mathrm{vdW}$ is the diameter of the smallest cylinder enclosing the molecule (including vdW radii), i.e.\ the limiting dimension for encapsulation into BNNTs. Uncertainties are on the order of $\pm 0.1$--$0.2$~nm.}
    \label{tab:vdw_dimensions_full}
    \begin{tabular}{l l c l}
    \toprule
    \textbf{Dye} & \textbf{Class} & \textbf{$d_\mathrm{vdW}$ (nm)} & \textbf{Comment} \\
    \midrule
    4T (quaterthiophene) & oligothiophene & 0.7       & Set by width of the thiophene plane + H vdW \\
    6T (sexithiophene)   & oligothiophene & 0.7       & Same cross section as 4T \\
    8T (octithiophene)   & oligothiophene & 0.7       & Same cross section as 4T \\
    \addlinespace
    Nile Red (NR)  & phenoxazine-like & 0.7--0.9 & Set by peripheral phenyl and carbonyl groups \\
    Nile Blue (NB) & phenoxazine      & 0.6--0.8 & Slightly more compact than NR \\
    \addlinespace
    Rhodamine B (RhodB)  & xanthene & 0.9--1.2 & Ethyl groups and twisted phenyl ring \\
    Rhodamine 6G (R6G)   & xanthene & 0.9--1.1 & Similar to RhodB, slightly more compact \\
    \addlinespace
    SQ3 & squaraine & 0.6--0.8        & Substituent-dependent \\
    SQA & squaraine & \(\lesssim\)1.4 & Wider due to donor groups \\
    SQB & squaraine & \(\lesssim\)1.1 & Similar range \\
    \bottomrule
    \end{tabular}
\end{table*}

\subsection{Ensemble investigations}

\subsubsection{Extinction and photoluminescence of free and encapsulated dyes}
\label{subsec:ensemble-spectra}

\begin{figure}[htbp]
    \centering
    \includegraphics[width=0.9\columnwidth]{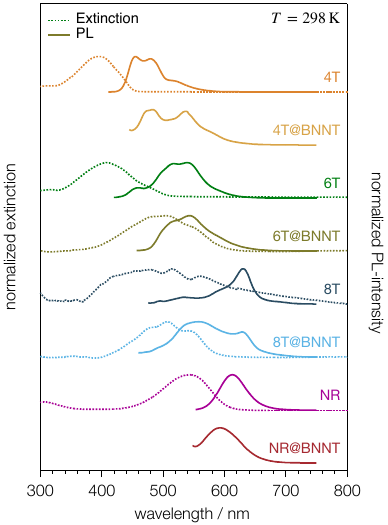}
    \caption{Normalized absorption (dashed) and photoluminescence (solid) spectra of representative dyes in solution and after encapsulation in BNNTs (X@BNNT) in DMF at room temperature. The comparison highlights systematic spectral changes upon encapsulation, including bathochromic shifts and modified vibronic structure, indicative of altered molecular conformation and intermolecular coupling under one-dimensional confinement. While spectral positions and line shapes vary between dyes, encapsulated species consistently exhibit signatures distinct from their solution counterparts. For 4T@BNNT and NR@BNNT, the transmission spectra show dye signatures that are too weak to be reliably separated from the scattering background.}
    \label{fig:results_single}
\end{figure}

Figure~\ref{fig:results_single} compares normalized extinction and PL spectra of the free dyes in DMF with those of the corresponding Dye@BNNT dispersions at room temperature. For all four systems, encapsulation produces reproducible spectral changes relative to the free dyes. Together with the single-particle measurements discussed later, these data show that emission from the BNNT samples reflects confinement-induced changes in molecular conformation, intermolecular coupling, and, for Nile Red, the local dielectric environment \cite{Allard2020,Jordan2023}. The ensemble spectra further suggest structurally heterogeneous encapsulated populations, which is plausible for BNNTs with a broad inner-diameter distribution and multiple possible endohedral packing motifs.

We first consider the three oligothiophenes in free solution in order to establish the spectral baseline of the series before turning to the corresponding BNNT-filled samples. Nile Red is discussed separately because its structureless spectra and pronounced solvatochromism require a different interpretive framework.

The solution spectra of the free oligothiophenes reflect the interplay of conjugation length and conformational disorder across the series. 4T provides the shortest-chain reference, with 6T showing only a modest red shift of the low-energy extinction onset by roughly 20~nm, whereas 8T exhibits a much broader spectral envelope extending nearly 200~nm further to the red. By contrast, the high-energy onset remains comparatively similar across all three dyes, shifting by only about 30~nm. This indicates that a substantial fraction of the ensemble still samples comparatively short effective conjugation lengths, consistent with large inter-ring twist angles that electronically fragment 6T and 8T into shorter segments, as expected from the relatively low rotational barriers about the inter-ring C--C single bonds \cite{Kolle2016,Torras2023}. The main spectral evolution across the series is therefore not a uniform red shift of the entire band, but a progressive broadening toward lower energies, most pronounced for 8T.

This increasingly asymmetric low-energy broadening is accompanied by only weak vibronic resolution at room temperature. Neither 4T nor 6T shows a clearly resolved vibronic progression in extinction, while 8T displays two weak features at 512 and 560~nm, corresponding to a spacing of approximately 1670~cm$^{-1}$, broadly consistent with coupling to a high-frequency backbone stretching mode of predominantly C=C character \cite{Bredas1996,Barford2013,Scholes2006}. Even in 8T, however, these features remain superimposed on a broad absorption envelope, consistent with a distribution of thermally accessible torsional conformations that modulate the effective conjugation length and electronic structure \cite{Kislitsyn2016,Park2018}. The pronounced and nearly linear red tail of 8T, tapering out at approximately 800~nm, is therefore most conservatively attributed to a broad population of low-energy conformers with enhanced local planarity and conjugation, although a minor contribution from weakly associated species cannot be excluded. \cite{Li2014}.

The emission spectra further sharpen this picture. For 4T and 6T, the fluorescence spectra show recognizable vibronic structure and a slightly narrower spectral distribution than the corresponding extinction spectra, consistent with relaxation toward a more planar and structurally better-defined emissive state on the \(S_1\) surface \cite{Bredas1996,Barford2013,Kolle2016}. By contrast, 8T exhibits a pronounced breakdown of mirror symmetry between extinction and emission, with only a very weak short-wavelength onset and a dominant low-energy emission band at 630~nm. This behavior is difficult to reconcile with simple Franck--Condon emission from the same state that dominates absorption and instead points primarily to pronounced excited-state relaxation and self-localization, while emission from a distinct low-energy emissive species cannot be excluded. The free-dye spectra thus already show that substantial spectral shifts, broadening, and changes in relative vibronic intensities can arise even without BNNT confinement from single-chain conformational relaxation and, in the case of 8T, possibly also from weak intermolecular association.

For 4T@BNNT, these free-dye considerations remain an essential baseline, but the observed spectral changes may now additionally reflect backbone planarization favored by confinement and host--guest interactions at the nanotube sidewalls, together with the associated increase in effective conjugation length \cite{Bredas1996,Barford2013,Yamashita12}, H-, J-, or mixed HJ-type excitonic coupling between neighboring chromophores \cite{Kasha1965,Spano2009,Spano2010}, and structural disorder within the confined ensemble \cite{Scholes2006,Clark2007}. The PL is not only red-shifted relative to the free dye but also substantially redistributed within the vibronic manifold. Compared with the free 4T spectrum, in which the two highest-energy emission features dominate and the intensity drops rapidly toward the red, the encapsulated spectrum shows a pronounced enhancement of lower-energy emission, including a strong band near 536~nm, a distinct shoulder around 568~nm, and a weak tail extending to about 640~nm. Because the extinction of 4T@BNNT cannot be separated reliably from the BNNT scattering background, the interpretation must remain cautious. Nevertheless, the PL line shape is more consistent with a confined aggregate showing appreciable H-like or mixed HJ character than with a simple bright J-aggregate, because the spectral weight is redistributed toward lower-energy vibronic emission rather than concentrated in the highest-energy band \cite{Kasha1965,Spano2009,Spano2010}. The remaining spectral breadth and red tail further point to disorder and a distribution of endohedral arrangements in the tube ensemble.

For 6T@BNNT, the dominant spectral change is the strong red-shift of the extinction spectrum, indicating a substantial lowering of the excitation energy in the confined ensemble, most plausibly due to confinement-induced planarization together with intermolecular coupling \cite{Bredas1996,Barford2013,Kasha1965,Spano2010}. On its own, however, this bathochromic shift does not justify assignment to a single idealized aggregate motif. The PL lacks the narrow, strongly 0--0-dominated emission expected for a simple bright J-aggregate, while the broad extinction and emission line shapes likewise argue against one well-defined H-type species. Instead, the spectra are most consistent with a structurally heterogeneous confined manifold in which planarization, intermolecular coupling, and disorder all contribute. This view is in line with earlier structural work on 6T inside BNNTs, which found predominantly H-like stacked arrangements and at most a minor contribution from strictly single-row J-type packing in ensemble measurements \cite{Allard2020}.

The encapsulated 8T system shows the most distinct response in the series. In extinction, the very broad and anomalous free-dye spectrum, including its long red tail extending to nearly 800~nm, collapses into a much narrower band with recognizable substructure and only a weak residual tail. The PL changes even more pronounced: whereas the free dye is dominated by a low-energy band around 630~nm, the encapsulated sample shows a broad higher-energy emission plateau with a maximum near 555~nm, while the 630~nm band survives as a secondary feature. Encapsulation therefore shifts the 8T emission profile away from the anomalously narrow low-energy manifold that dominates in solution and toward a spectrally broader emissive ensemble. This suggests that confinement suppresses access to the most strongly relaxed states that dominate in the free sample and instead favors a narrower set of more rigid conformations. At the same time, the persistence of the 630~nm band and the remaining spectral breadth indicate that the encapsulated ensemble remains heterogeneous. Because the free 8T reference is already spectroscopically complex, the comparison is less direct than for 4T or 6T. The encapsulated spectrum is therefore best interpreted primarily in terms of rigidification and conformational selection, with intermolecular coupling likely present but not cleanly resolved spectroscopically.

Overall, the oligothiophene series shows that BNNT encapsulation reshapes the optical response in a strongly length-dependent manner. For 8T, encapsulation suppresses the anomalous low-energy manifold that dominates the free reference and instead favors a more restricted emissive ensemble at higher energy. For 6T, the dominant effect is a pronounced red-shift of the extinction spectrum, while the emission remains broadly similar in overall shape and does not support assignment to a simple bright J-aggregate. For 4T, confinement primarily redistributes emission intensity toward the lower-energy part of the vibronic manifold and enhances the red tail. BNNT confinement therefore does not impose a single universal spectroscopic response, but rather changes the relative contributions of intramolecular rigidification, conformational selection, and intermolecular coupling in a manner that depends sensitively on oligomer length and endohedral packing. The measured spectra should accordingly be understood as ensemble averages over structurally heterogeneous confined states and packing motifs \cite{Badon2023}.

\begin{table*}[h]
\centering
\caption{Comparison of PL quantum yields, weighted average fluorescence lifetimes, and effective radiative and non-radiative decay rates for dyes in DMF and after encapsulation inside BNNTs. The rates were calculated from \(k_\mathrm{r}=\Phi_{\rm PL}/\tau_\mathrm{PL}\) and \(k_\mathrm{nr}=(1-\Phi_{\rm PL})/\tau_\mathrm{PL}\).}
\label{tab:rates-oligothiophenes}
\newcolumntype{C}{>{\centering\arraybackslash}X}
\begin{tabularx}{\textwidth}{l CCCCC}
\toprule
\textbf{System} & $\Phi_{\rm PL}$ & $\tau_\mathrm{PL}$ (ns) & $k_\mathrm{tot}$ (ns$^{-1}$) & $k_\mathrm{r}$ (ns$^{-1}$) & $k_\mathrm{nr}$ (ns$^{-1}$) \\
\midrule
4T in DMF         & 0.08 & 0.98  & 1.02 & 0.08 & 0.94 \\
4T@BNNT in DMF    & 0.02 & 1.08  & 0.93 & 0.02 & 0.91 \\ \addlinespace
6T in DMF         & 0.57 & 0.88 & 1.13  & 0.65 & 0.49 \\
6T@BNNT in DMF    & 0.06 & 0.81 & 1.23  & 0.07 & 1.16 \\ \addlinespace
8T in DMF         & 0.06 & 0.76  & 1.31 & 0.08 & 1.24 \\
8T@BNNT in DMF    & 0.05 & 0.79 & 1.26 & 0.06 & 1.20 \\ \addlinespace
\bottomrule
\end{tabularx}
\end{table*}

For NR@BNNT, the extinction again could not be separated cleanly from the BNNT scattering background, but the PL maximum shifts hypsochromically from about \SI{615}{nm} for the free dye to \SI{595}{nm} after encapsulation. In contrast to the oligothiophenes, this blue shift is most plausibly attributed to the well-known solvatochromic response of Nile Red's intramolecular charge-transfer (ICT) excited state rather than to excitonic aggregate formation. Because the excited-state dipole moment is substantially larger than that of the ground state, the emission energy is highly sensitive to dielectric stabilization of the ICT manifold by the surrounding medium~\cite{Reichardt2011,Greenspan1985,Hess2003,Gajo2024}. In polar DMF, the ICT state is stabilized efficiently and the emission is correspondingly red-shifted. A transfer into the weakly polarizable interior of the BNNT raises the relaxed excited-state energy and shifts the PL to the blue. An empirical Nile Red emission--permittivity calibration~\cite{Hess2003,Fuhl2024} places the effective dielectric constant of the confined environment in the range $\varepsilon \approx 4$--$5$, well below that of bulk DMF ($\varepsilon \approx 38$). We caution that aggregation effects cannot be excluded from the PL shift alone and may contribute additionally to line-shape and intensity changes. For Nile Red, however, the dominant origin of the hypsochromic shift appears to be the change in local dielectric environment rather than strong excitonic coupling. In this sense, the BNNT acts here predominantly as a low-permittivity confining cavity rather than as a template for the strongly exciton-coupled aggregates inferred for the oligothiophenes.

Across the dye series, the ensemble data reveal clear and dye-specific spectral signatures of BNNT encapsulation. For the oligothiophenes, the observed changes point to a length-dependent interplay of conformational rigidification, conformational selection, and intermolecular coupling under one-dimensional confinement. For Nile Red, by contrast, the dominant effect is the dielectric destabilization of a solvatochromic ICT excited state in the lower-permittivity BNNT interior. These findings show that BNNT encapsulation does not merely perturb molecular spectra in a nonspecific way, but can direct the photophysical response along distinct mechanistic pathways. The contrast between oligothiophene-like confinement-modified excitonic landscapes and the dielectric tuning of Nile Red's ICT emission highlights the mechanistic versatility of BNNT nanoconfinement across different dye classes.

\subsubsection{Oligothiophene PL quantum yields and lifetimes}
\label{subsec:qy-lifetime-comparison}

\begin{figure}[htbp]
    \centering
    \includegraphics[width=0.9\columnwidth]{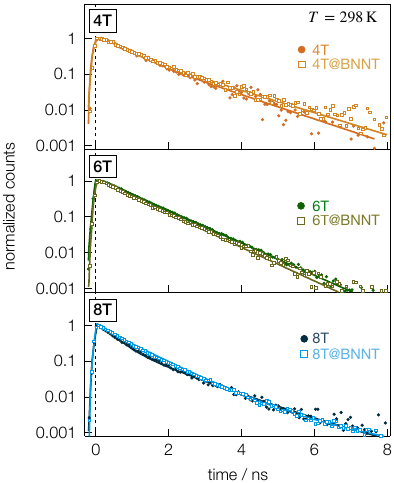}
    \caption{Time-correlated single-photon counting (TCSPC) decay curves of oligothiophenes (4T, 6T, 8T) in DMF and after encapsulation in BNNTs (X@BNNT) at room temperature (\(T = 298\)~K). The data are shown as normalized intensities on a semi-logarithmic scale. Symbols represent the experimental data, while solid lines correspond to reconvolution fits.}
    \label{fig:TCSPC}
\end{figure}

To assess how encapsulation inside BNNTs affects excited-state relaxation, we compare absolute photoluminescence (PL) quantum yields and fluorescence lifetimes of the free dyes in DMF with those of the corresponding Dye@BNNT dispersions. The TCSPC decays shown in Figure~\ref{fig:TCSPC} were described by multiexponential reconvolution fits. For the quantitative comparison below, we use weighted average lifetimes, \(\tau_\mathrm{PL}\), obtained from the fitted amplitudes and lifetime components according to \(\tau_\mathrm{PL}=\sum_i a_i\tau_i^2/\sum_i a_i\tau_i\). Combined with the ensemble PL quantum yield \(\Phi_{\rm PL}\), these values define effective radiative and non-radiative rates via \(k_\mathrm{r}=\Phi_{\rm PL}/\tau_\mathrm{PL}\) and \(k_\mathrm{nr}=(1-\Phi_{\rm PL})/\tau_\mathrm{PL}\). The resulting values for 4T, 6T, and 8T are summarized in Table~\ref{tab:rates-oligothiophenes}.

Before turning to the encapsulated systems, it is useful to place our solution measurements of 4T, 6T, and 8T in DMF in the context of earlier solution-phase studies of $\alpha$-oligothiophene photophysics. The general trends observed here --- a low PL quantum yield for 4T, a pronounced maximum for 6T, and a renewed decrease for 8T, accompanied by sub-nanosecond fluorescence lifetimes throughout the series --- are broadly consistent with the picture established in earlier solution-phase work on $n$T photophysics \cite{Becker1996,Rentsch1999,Lap1997}. Our measured value for 4T and the high fluorescence efficiency found for 6T are broadly consistent with reported solution-phase values. The occurrence of a fluorescence-efficiency maximum near 6T is well established and is generally rationalized by a length-dependent crossover between ISC and IC as the dominant non-radiative channels, with both channels modulated by inter-ring torsional dynamics on the $S_1$ surface \cite{Becker1996,Rentsch1999,Beljonne1996,Kolle2016}. Our low 8T quantum yield is likewise consistent with the additional conformational relaxation, disorder, and possible weak association expected for longer oligothiophenes \cite{Park2018,Li2014}. Our solution data thus broadly reproduce the established photophysical hierarchy of the $n$T series and provide a baseline against which the encapsulation-induced changes can be interpreted.

The kinetic consequences of BNNT encapsulation are summarized in Table~\ref{tab:rates-oligothiophenes}. A useful empirical constraint is that the total decay rate $k_\mathrm{tot}$ varies only modestly across all six systems, lying between $0.93$ and $1.31$~ns$^{-1}$, so that encapsulation changes the overall excited-state decay rate by no more than about $30\,\%$. What changes substantially is the effective partitioning between radiative and non-radiative decay, in a chain-length-dependent way.

The factor-of-four to factor-of-ten reductions of $k_\mathrm{r}$ seen in Table~\ref{tab:rates-oligothiophenes} cannot be attributed to the change in dielectric environment alone, because standard refractive-index and local-field corrections would be expected to produce changes of order 1, not the more pronounced suppressions observed here. Standard refractive-index and local-field corrections would be expected to produce changes of at most order unity, not the suppressions observed here. The decrease in effective $k_\mathrm{r}$ therefore most likely reflects changes in the emissive state or emitting population rather than a simple dielectric-environment effect. At the same time, neither a shortening of $\tau_\mathrm{PL}$ nor an enhancement of $k_\mathrm{r}$ is observed for any encapsulated system, strongly disfavoring a bright superradiant J-aggregate assignment. Together with the spectral analysis (Section~\ref{subsec:ensemble-spectra}), the kinetic data are most consistent with structurally heterogeneous confined ensembles with H-like or mixed HJ-type coupling \cite{Allard2020,Spano2010}.

For 4T, encapsulation reduces \(k_\mathrm{r}\) by roughly a factor of four, from \(0.08\) to \(0.02\)~ns\(^{-1}\), while \(k_\mathrm{nr}\) remains essentially unchanged (\(0.94\) to \(0.91\)~ns\(^{-1}\)). The literature assigns the inefficient fluorescence of short oligothiophenes such as 4T largely to ISC-related decay, governed by intramolecular spin--orbit couplings and singlet--triplet energy gaps \cite{Kolle2016,Beljonne1996}. The near-invariance of \(k_\mathrm{nr}\) suggests that these non-radiative channels are not strongly perturbed at the ensemble level. By contrast, the selective suppression of effective \(k_\mathrm{r}\) is consistent with an environment-induced change in the emissive population, in line with the H-like or mixed HJ-type spectral response established for 4T@BNNT (Section~\ref{subsec:ensemble-spectra}).

For 6T, the kinetic response is qualitatively different and the most pronounced of the three systems. Encapsulation reduces \(k_\mathrm{r}\) by nearly an order of magnitude, from \(0.65\) to \(0.07\)~ns\(^{-1}\), while \(k_\mathrm{nr}\) rises from \(0.49\) to \(1.16\)~ns\(^{-1}\). Free 6T lies near the radiatively favored maximum of the \(n\)T series, with its high \(\Phi_\mathrm{PL}\) reflecting strong oscillator strength and comparatively inefficient ISC \cite{Kolle2016}. Encapsulation shifts this balance strongly. The spectra indicate confinement-induced planarization together with intermolecular coupling (Section~\ref{subsec:ensemble-spectra}), but the suppression of \(k_\mathrm{r}\) shows that the emissive population is not a bright monomer-like or superradiant J-like state. The concomitant rise in \(k_\mathrm{nr}\) suggests that additional or previously inefficient non-radiative channels become competitive, plausibly through energy migration to lower-lying disordered sites within the heterogeneous confined ensemble. Of the three oligothiophenes, 6T shows the largest kinetic response because the free-solution balance is initially most favorable to radiative decay. The partial compensation between reduced \(k_\mathrm{r}\) and increased \(k_\mathrm{nr}\) explains why \(k_\mathrm{tot}\) nevertheless changes only modestly.

For 8T, by contrast, the kinetic data show no significant kinetic response to encapsulation: \(k_\mathrm{r}\) changes only marginally from \(0.08\) to \(0.06\)~ns\(^{-1}\), \(k_\mathrm{nr}\) from \(1.24\) to \(1.20\)~ns\(^{-1}\), and \(\tau_\mathrm{PL}\) from \(0.76\) to \(0.79\)~ns. This invariance is striking given that the spectral response of 8T is among the most pronounced in the series. The effective non-radiative balance associated with the relaxed 8T emissive ensemble is therefore largely preserved under confinement, even as the conformational distribution is substantially restricted \cite{Beljonne1996,Lap1997,Park2018}. Of the three oligothiophenes, 8T is the system in which BNNT confinement most clearly restricts the conformational ensemble without substantially altering the decay kinetics.

A comparable indication of ensemble heterogeneity emerges from the time-resolved data on NR@BNNT, despite the very different photophysics of the chromophore. Free Nile Red in DMF decays monoexponentially with $\tau \approx \SI{4.1}{ns}$, in agreement with literature values for the dye in moderately polar solvents~\cite{Hess2003,Fuhl2024}, and the lifetime is concentration-independent up to several hundred~\si{\micro M}. Upon encapsulation, the decay becomes distinctly biexponential with an amplitude-weighted mean lifetime of approximately \SI{0.8}{ns}~\cite{Fuhl2024}. The shortening is qualitatively consistent with destabilization of the polarity-stabilized ICT emissive state in the lower-permittivity BNNT interior~\cite{Hess2003,Gajo2024}, although the present data do not separate radiative from non-radiative contributions to the lifetime change. The pronounced biexponentiality, absent in solution, suggests that the encapsulated ensemble samples more than one local environment, consistent with the structural heterogeneity inferred for the oligothiophene hybrids.

Taken together, the kinetic data on the oligothiophene series reveal a length-specific pattern that the spectral analysis alone could not resolve. BNNT encapsulation keeps the total excited-state decay rate within a relatively narrow range across the series but redistributes the balance between radiative and non-radiative decay in a way that reflects the dominant free-dye relaxation balance. The picture that emerges is one of length-specific, mechanism-selective perturbation, in which confinement changes the effective radiative and non-radiative contributions to different extents.

\subsection{Single-particle spectroscopy of oligothiophene-filled BNNTs}
\label{subsec:single-particle-pl-polarization-oligothiophenes}

Polarization-resolved PL microscopy on individual oligothiophene-filled BNNTs provides direct evidence that encapsulated emitters can adopt strongly anisotropic arrangements within single nanotubes. At the same time, the measurements reveal pronounced tube-to-tube variability in both polarization degree and apparent polarization angle, indicating that confinement does not impose one unique internal geometry but rather a distribution of ordered configurations.

\begin{figure}[htbp]
    \centering
    \includegraphics[width=0.9\columnwidth]{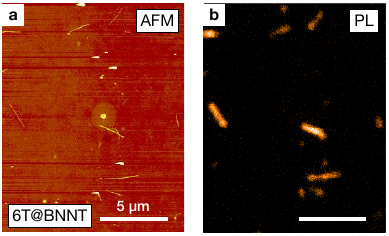}
    \caption{Correlated AFM and PL images of a representative spin-coated 6T@BNNT sample area on the substrate. \textbf{a,} AFM image showing about a dozen nanotube-like objects (\qty{5}{\micro\meter} scale bar). \textbf{b,} PL image of exactly the same area. Most AFM-visible objects also appear as PL-active rods or, for very short nanotubes, as diffraction-limited spots. The comparison provides a qualitative impression that a substantial fraction of the deposited nanotubes is emissive after purification, although it is not intended as a quantitative determination of the filling yield.}
    \label{fig:fig5_afm_pl_overview}
\end{figure}

As a qualitative starting point, correlated AFM and PL imaging suggests that a substantial fraction of the deposited nanotubes remains emissive after purification. Figure~\ref{fig:fig5_afm_pl_overview} compares an AFM image of a representative 6T@BNNT sample area with the corresponding PL image of exactly the same region. Most rod-like objects in the AFM image, consistent with individual nanotubes or small nanotube bundles, also appear in PL as elongated emitters or, for very short nanotubes, as diffraction-limited spots. This comparison is not intended as a rigorous determination of the filling fraction, since overlap of objects, finite optical resolution, bundling, and selection bias toward brighter emitters complicate exact counting. Nevertheless, the strong correspondence between AFM-visible nanotubes and PL-active objects supports the conclusion that a sizable fraction of the deposited BNNTs remains optically active after the washing procedure, consistent with successful dye encapsulation.

\begin{figure}[htbp]
    \centering
    \includegraphics[width=0.9\columnwidth]{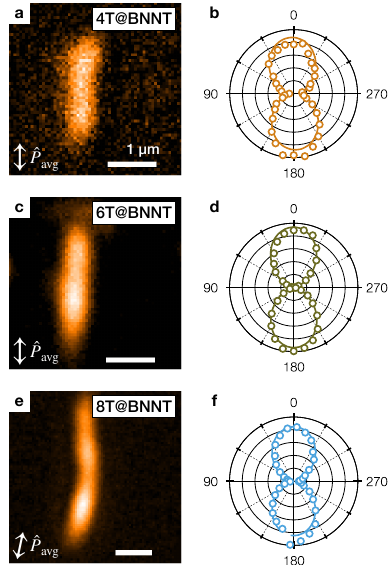}
    \caption{Representative polarization-resolved PL data for individual oligothiophene-filled BNNTs. \textbf{a,c,e,} False-color PL images of representative 4T@BNNT, 6T@BNNT, and 8T@BNNT nanotubes. The 4T@BNNT and 6T@BNNT nanotubes are about \qty{2}{\micro\meter} long, and the 8T@BNNT nanotube about \qty{3}{\micro\meter}. All three objects show emission distributed over most of the visible tube length, with moderate intensity variations along the tube. Double-headed arrows indicate the fitted polarization directions \(\hat P_\mathrm{avg}\). \textbf{b,d,f,} Corresponding polarization dependences of the integrated PL intensity. Solid lines show \(\cos^{2}(\vartheta)\)-type fits to the data, with \(\vartheta\) being the angle with respect to the tube axis. All three examples exhibit polarization degrees close to unity. The fitted average polarization axis is essentially parallel to the tube axis for 4T@BNNT and 6T@BNNT, while the 8T@BNNT example shows a small apparent offset of about \qty{12}{\degree}, possibly influenced by the slight bend of the nanotube.}
    \label{fig:fig6_representative_polarized_tubes}
\end{figure}

Representative examples of individual 4T@BNNT, 6T@BNNT, and 8T@BNNT nanotubes are shown in Figure~\ref{fig:fig6_representative_polarized_tubes}. In all three cases, the PL extends over essentially the full visible tube length rather than being confined to a single localized hotspot. Moderate intensity variations along the tube are common; in the examples shown here, the local intensity varies by up to roughly 40\% along the tube axis, which may reflect non-uniform filling, local packing variations, or local quenching sites. The corresponding polarization-dependent PL traces are well described by a \(\cos^{2}\)-type dependence. In the following, the degree of polarization is quantified as \(P = (I_{\max}-I_{\min})/(I_{\max}+I_{\min})\), where \(I_{\max}\) and \(I_{\min}\) denote the fitted maximum and minimum emission intensities, respectively. The large \(P\) values obtained for these representative nanotubes, all close to unity, show that highly polarized emission is not an isolated exception but can occur in all three oligothiophene systems. For the 4T@BNNT and 6T@BNNT examples, the polarization axis is essentially parallel to the tube axis. In the 8T@BNNT example, the apparent polarization direction is tilted by about \qty{12}{\degree}; given that the nanotube itself is slightly curved, part of this deviation may reflect the tube geometry rather than a true intrinsic tilt of the emitting dipoles. These examples are intentionally selected to illustrate strongly polarized emission, while the full distributions of polarization degree and apparent polarization angle are analyzed below.

\begin{figure}[htbp]
    \centering
    \includegraphics[width=0.9\columnwidth]{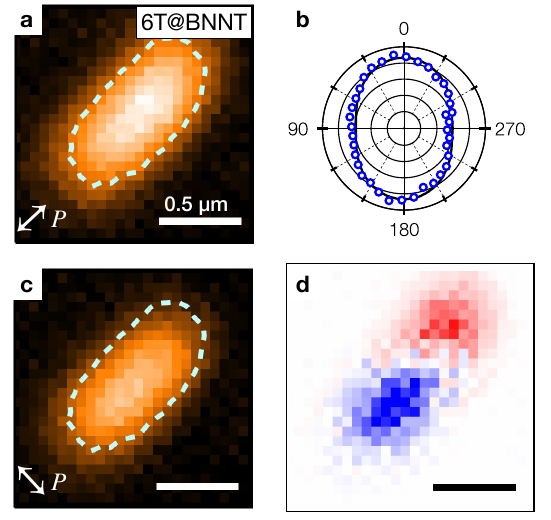}
    \caption{Example of spatially inhomogeneous polarization within a single 6T@BNNT nanotube. \textbf{a,} PL image \(I_{\parallel}\) recorded at the analyzer angle giving the maximum nanotube-integrated intensity. The dashed contour indicates the region where the intensity exceeds 50\% of the peak value. \textbf{b,} Polar plot of the nanotube-integrated PL intensity, yielding a polarization degree of \(P=\num{0.32}\). \textbf{c,} PL image \(I_{\perp}\) of the same nanotube recorded with the analyzer rotated by \qty{90}{\degree} relative to panel \textbf{a}. The dashed contour from panel \textbf{a} is shown for reference. \textbf{d,} Signed spatial contrast map derived from \(I_{\parallel}\) and \(I_{\perp}\) using the tube-integrated polarization ratio as reference. Blue indicates locally stronger alignment with the dominant polarization direction, red weaker alignment, and white corresponds to the tube-average polarization balance.}
    \label{fig:fig7_inhomogeneous_single_tube}
\end{figure}

At the same time, not all nanotubes are as uniformly polarized as the examples in Figure~\ref{fig:fig6_representative_polarized_tubes}. Figure~\ref{fig:fig7_inhomogeneous_single_tube} shows a roughly \qty{1}{\micro\meter} long 6T@BNNT nanotube with a much smaller tube-integrated polarization degree of only \(P=\num{0.32}\). The PL image recorded at the analyzer angle of maximum intensity still shows a clearly elongated emitter, but the orthogonal analyzer setting reveals substantial brightness changes along the tube. In particular, the upper-right section emits relatively weakly in the orthogonal polarization channel, whereas the lower-left section remains comparatively bright. To visualize these local differences more clearly, Figure~\ref{fig:fig7_inhomogeneous_single_tube}d shows a signed contrast map constructed from the two polarization-resolved images, using the tube-integrated intensity ratio to define the relative weighting of the parallel and orthogonal channels. This representation highlights local deviations from the average polarization balance of the nanotube: blue regions are locally more strongly aligned with the dominant polarization direction, whereas red regions are less strongly aligned. The map therefore shows that the local polarization balance is not spatially uniform along the nanotube. The tube-integrated polarization degree consequently averages over locally distinct polarization characters, and locally reduced contrast can reflect either a different local polarization direction or genuinely lower local order. Such intratube heterogeneity is consistent with local variations in filling, packing, or quenching within a single nanotube.

\begin{figure}[htbp]
    \centering
    \includegraphics[width=0.9\columnwidth]{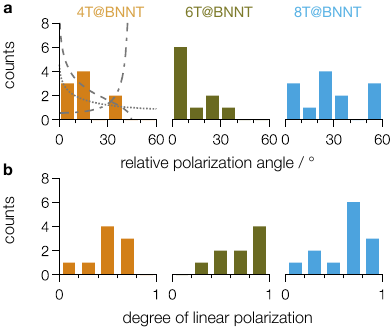}
    \caption{Histograms of apparent polarization angles and polarization degrees for oligothiophene-filled BNNTs. \textbf{a,} Histograms of the apparent in-plane angle \(\vartheta\) between the fitted polarization axis and the nanotube axis for 4T@BNNT (\(n=9\)), 6T@BNNT (\(n=9\)), and 8T@BNNT (\(n=13\)). In the 4T@BNNT panel, the short-dashed, long-dashed, and dash-dotted gray lines indicate simulated projected distributions expected for isotropic \((\theta,\phi)\) ensembles with \(\theta_{\max}=\qty{45}{\degree}\), with \(\theta_{\max}=\qty{90}{\degree}\), and for a fixed \(\theta=\qty{45}{\degree}\) with isotropic \(\phi\), respectively. \textbf{b,} Corresponding histograms of the polarization degree \(P\) for the same nanotube sets. All three systems are skewed toward higher polarization degrees, with comparatively few nanotubes in the lowest \(P\) bin and many objects in the range \(P=\numrange{0.5}{1.0}\). The limited statistics do not justify a detailed comparison to specific orientational models, but they clearly show that strongly polarized emission is common in all three oligothiophene@BNNT systems.}
    \label{fig:fig8_histograms_angles_polarization}
\end{figure}

The interpretation of apparent polarization angles requires some geometrical care. Let \(\theta\) denote the true polar angle of the emitting transition dipole moment with respect to the nanotube axis and \(\phi\) the azimuthal angle around that axis. In normal-emission microscopy, however, only the in-plane projection of this orientation is experimentally accessible. We denote the angle between the projected polarization axis and the projected tube axis by \(\vartheta\). For a nanotube lying in the substrate plane, the projection geometry gives \(\tan \vartheta = \tan \theta\,|\cos\phi|\), so that \(0 \le \vartheta \le \theta\). Consequently, even if all emitters in a nanotube were perfectly co-aligned at one fixed nonzero angle \(\theta\), the experimentally observed in-plane polarization angle could still assume any smaller value depending on \(\phi\). The measured angle distribution therefore cannot be interpreted directly as the true three-dimensional tilt distribution. This projection effect is illustrated in the left panel of Figure~\ref{fig:fig8_histograms_angles_polarization}a, where the dash-dotted gray guide curve corresponds to the projected distribution expected for a fixed \(\theta=\qty{45}{\degree}\) and isotropic \(\phi\), and it must be borne in mind when comparing the measured angle histograms across the three dye systems.

With this geometrical limitation in mind, the measured distributions of apparent polarization angles \(\vartheta\) in Figure~\ref{fig:fig8_histograms_angles_polarization} are broad for all three oligothiophene systems. In each case, some nanotubes fall into the \qtyrange{0}{10}{\degree} bin, showing that nearly axial projected polarization is present in all samples. At the same time, larger apparent angles are also observed, reaching roughly \qtyrange{30}{40}{\degree} for 4T@BNNT and 6T@BNNT and up to about \qtyrange{50}{60}{\degree} for 8T@BNNT. The gray guide curves included for 4T@BNNT illustrate three simple projected model distributions, namely isotropic \((\theta,\phi)\) ensembles with \(\theta_{\max}=\qty{45}{\degree}\), isotropic \((\theta,\phi)\) ensembles with \(\theta_{\max}=\qty{90}{\degree}\), and the projected distribution for fixed \(\theta=\qty{45}{\degree}\) combined with isotropic \(\phi\). Given the limited statistics (\(n=9\), \(9\), and \(13\) objects for 4T@BNNT, 6T@BNNT, and 8T@BNNT, respectively), the present data do not allow a clear discrimination between such model distributions. They do, however, show that the individual nanotubes span a broad range of projected orientations, with mean apparent angles of about \qty{16}{\degree}, \qty{13}{\degree}, and \qty{28}{\degree} for 4T@BNNT, 6T@BNNT, and 8T@BNNT, respectively. The distinctly larger mean value for 8T@BNNT is consistent with the broader orientational variability already suggested by its single-tube and ensemble spectral behavior, although the present dataset is too limited to support a more specific structural assignment.

The histograms of polarization degree in Figure~\ref{fig:fig8_histograms_angles_polarization} are particularly informative in a qualitative sense. For all three oligothiophene systems, the distributions are skewed toward higher \(P\) values. Only a few nanotubes fall into the \numrange{0.0}{0.2} bin, whereas roughly half of the analyzed objects lie between \(P=\num{0.5}\) and \(1.0\). Thus, although strongly polarized emission is not universal, it is clearly common across all three systems. Together with the broad distribution of apparent polarization angles, this shows that many nanotubes contain emitting populations with substantial orientational order, even though the projected polarization direction varies from tube to tube.

The single-particle polarization data sharpen the picture obtained from ensemble spectroscopy, PL quantum yields, and fluorescence lifetimes. Strong polarization from an individual nanotube does not by itself identify the underlying coupling geometry, but it does show that the emitting populations are often orientationally well ordered within the tube. The ensemble measurements had already shown that encapsulation does not produce one unique aggregate type, but rather a heterogeneous distribution of confined states shaped by planarization, intermolecular coupling, and conformational selection. The polarization microscopy now adds that the emission from many of these confined ensembles is nevertheless strongly anisotropic at the level of individual nanotubes. For 4T and 6T, this is consistent with the earlier conclusion that confinement promotes ordered but spectroscopically mixed arrangements rather than one unique packing motif, in line with previous reports of diameter-dependent packing and structural heterogeneity in BNNT hosts \cite{Badon2023}. For 8T, the broader spread in apparent angles and polarization degrees suggests a less uniform orientational organization, consistent with the greater conformational and spectroscopic heterogeneity already inferred from the ensemble data.

Overall, the polarization-resolved single-particle measurements support three main conclusions. First, all three oligothiophene@BNNT systems, 4T@BNNT, 6T@BNNT, and 8T@BNNT, can exhibit strongly polarized emission from individual nanotubes. Second, the internal orientational order varies substantially from tube to tube and even along a single nanotube. Third, the broad apparent angle distributions must be interpreted with geometrical caution, because normal-emission microscopy probes only the projected polarization direction rather than the full three-dimensional dipole orientation. Within the present statistics, the most robust conclusion is therefore that BNNT confinement frequently generates elongated emitting populations with partial to high orientational order, while still allowing substantial structural heterogeneity between different nanotubes.

\section{Conclusions}
Wet-chemical postsynthetic filling of BNNTs with ten dyes under a common preparation and purification protocol reveals that successful encapsulation is strongly guest-specific and cannot be predicted from steric fit alone. Clear spectroscopic signatures of stable encapsulation were obtained for 4T, 6T, 8T, and Nile Red, while Nile Blue, Rhodamine B, Rhodamine 6G, and three squaraines failed to yield stable host--guest systems under identical conditions. The observed selectivity reflects distinct limiting factors for different dye classes, including solvation and aggregation behavior for the squaraines and chemical stability during filling and washing for the rhodamines, rather than a single dominant criterion \cite{Allard2020,Badon2023,Jordan2023}.

For the successfully encapsulated systems, BNNT confinement produces guest-specific optical responses rather than a universal confinement signature. In the oligothiophene series, encapsulation reshapes both spectra and excited-state dynamics in a strongly oligomer--length-dependent manner, consistent with different balances of rigidification, conformational selection, and intermolecular coupling. Quantitatively, the total excited-state decay rate of the oligothiophenes is approximately conserved under encapsulation, while the partitioning between radiative and non-radiative channels shifts in a length- and mechanism-specific way, with 6T showing the strongest redistribution. This kinetic invariance, together with the absence of fluorescence-lifetime shortening across the series, disfavors bright J-aggregate assignments and is more consistent with confinement that restricts the conformational ensemble without substantially altering the decay kinetics. Nile Red provides a complementary case in which the dominant response is more naturally attributed to dielectric confinement of a solvatochromic ICT state than to strongly exciton-coupled aggregates \cite{Gajo2024}. Single-particle polarization microscopy further shows that micrometer-long oligothiophene-filled BNNTs frequently generate strongly polarized emission from individual nanotubes, even within ensembles that remain structurally heterogeneous between tubes \cite{Allard2020,Badon2023}.

More generally, these findings highlight an important distinction between BNNT and SWCNT host systems. Encapsulation inside SWCNTs has long been used to create one-dimensional molecular assemblies with controlled packing and direct host--guest electronic interactions, but the electronically active carbon nanotube often participates directly through absorption or excitation-energy transfer \cite{Smith1998,Forel2022}. BNNTs, by contrast, provide a wide-bandgap nanoconfined host in which molecular alignment, dielectric effects, and guest--guest coupling can be probed more directly through the guest response \cite{Allard2020}. Dye-filled BNNTs thus provide a particularly clean platform for disentangling how one-dimensional confinement modifies molecular excitations, packing, and emission anisotropy in the absence of dominant electronic contributions from the host itself.

These results extend earlier BNNT dye studies from selected model systems toward a more comparative and physically informed picture that links filling outcomes to dye-specific physicochemical properties \cite{Juergensen2025}. Further progress will depend on tighter structural control of the host nanotubes and more direct structure--property correlation. In particular, diameter-enriched BNNT samples with better discrimination of cavity size and wall structure should make it easier to separate host-diameter-dependent effects from effects arising from intrinsic packing motifs. Equally important will be correlated single-nanotube measurements combining optical spectroscopy with structural characterization, ideally with high spatial, spectral, and temporal resolution to resolve intratube heterogeneity and internal relaxation or energy-transfer processes. Together, such advances would move the field toward a more predictive understanding of how specific nanoconfined environments shape guest organization and photophysical response.

\section*{Acknowledgements}
The authors would like to thank C.~Kingston and his team as well as M. McArthur at the National Research Council Canada for a generous donation of BNNT materials. This work was supported by the Canada Research Chair, the Canada Foundation for Innovation, and the Natural Sciences and Engineering Research Council of Canada (NSERC grant number RGPIN-2025-05339). R.M.\ is a member of the RQMP network of the Fonds de Recherche du Québec (FRQ) (\url{https://doi.org/10.69777/309032}) and of the CQMF network of the FRQ (\url{https://doi.org/10.69777/341666}).

\appendix
\section{Declaration of generative AI and AI-assisted technologies in the manuscript preparation process}

During the preparation of this work, the authors used generative AI tools to improve the structure and presentation of the manuscript and to assist with LaTeX formatting. No AI tools were used to generate or analyze scientific data or to derive scientific conclusions. The authors take full responsibility for the final content of the manuscript.
\label{app1}






\bibliography{bib}
\end{document}